\documentclass[twocolumn]{aa}

\usepackage{graphicx}
\usepackage{txfonts}
\usepackage{epsfig}
\usepackage{amsbsy}
\usepackage{amssymb}
\usepackage{amsmath}
\usepackage{latexsym}
\usepackage{natbib}
\usepackage{verbatim}
\usepackage{bm}
\usepackage{tabularx}
\usepackage{soul}
\usepackage{xcolor}

\begin{document}

% \title{{Unsuitability of CRD and PRD--AA approximations for chromospheric 
% magnetic field diagnostics via\\ 
% forward-scattering Hanle effect}}

\title{
Accurate PRD modeling of the forward-scattering Hanle effect\\
in the chromospheric Ca~{\sc i} 4227\,{\AA} line}

\titlerunning{Forward-scattering Hanle effect in Ca~{\sc i} 4227}
\authorrunning{Belluzzi et al.}

\author{
	Luca Belluzzi\inst{1,2,3}
	\and 
	Simone Riva\inst{2}
	\and
	Gioele Janett\inst{1,2}
    \and
    Nuno Guerreiro\inst{1,2}
    \and
    Fabio Riva\inst{1,2}
	\and
	Pietro Benedusi\inst{2,4}
	\and
    \\
    Tanaus\'u del Pino Alem\'an\inst{5,6}
	\and
	Ernest Alsina Ballester\inst{5,6}
    \and
    Javier Trujillo Bueno\inst{5,6,7}
	\and
	Ji\v{r}\'i \v{S}t\v{e}p\'an\inst{8}
}

\institute{
	Istituto ricerche solari Aldo e Cele Dacc\`o (IRSOL), Faculty of informatics, Universit\`a della Svizzera italiana, Locarno, Switzerland
    \and
    Euler Institute, Universit\`a della Svizzera italiana, Lugano, Switzerland
    \and
    Institut f\"ur Sonnenphysik (KIS), Freiburg i.~Br., Germany
    \and   
	Simula Research Laboratory, Oslo, Norway
	\and
	Instituto de Astrof\'isica de Canarias, La Laguna, Tenerife, Spain
	\and
	Departamento de Astrof\'isica, Universidad de La Laguna, La Laguna, Tenerife, Spain
	\and
	Consejo Superior de Investigaciones Cient\'ificas, Spain
	\and
	Astronomical Institute ASCR, Ond\v{r}ejov, Czech Republic\\
	\email{luca.belluzzi@irsol.usi.ch}
}

\newcommand{\corange}[1]{\textcolor{orange}{#1}}
\newcommand{\cred}[1]{\textcolor{red}{#1}}
\newcommand{\MyNewEdit}[1]{\textcolor{green}{#1}}
\newcommand{\eab}[1]{\textcolor{teal}{#1}}

\abstract
{
%Precious information on chromospheric magnetic fields is encoded in the disk center linear polarization signals
%that they can induce in the core of strong resonance lines via the forward-scattering Hanle effect.
Measurable linear scattering polarization signals have been predicted and detected at the solar disk center
in the core of
% both photospheric and
chromospheric lines.
These forward-scattering polarization signals, which are of high interest for magnetic field diagnostics,
have always been modeled either under the assumption of complete frequency redistribution (CRD), or taking partial frequency 
redistribution (PRD) effects into account under the angle-averaged (AA) approximation.
}
%
% {\textcolor{orange}{
% In this work, \emph{for the first time(?)}, these signals are modeled by 
% solving the radiative transfer problem for polarized radiation, out of local 
% thermodynamic equilibrium, in one-dimensional semi-empirical models of the 
% solar atmosphere, accounting for PRD effects in the most general, 
% angle-dependent setting.}}
{
This work aims at assessing the suitability of the CRD and PRD--AA approximations for %the modeling of the forward-scattering Hanle effect
modeling the forward-scattering polarization signals produced by the presence of an inclined magnetic field, the so-called forward-scattering Hanle effect,
in the chromospheric Ca~{\sc i} 4227~{\AA} line.
}
%
% {\textcolor{orange}{
% We focus on the chromospheric Ca~{\sc i} line at 4227\,{\AA}, which we model considering a two-level atom. 
% The problem is linearized by calculating a priori the population of the lower 
% level,
% and it is solved through a novel solution strategy based on 
% preconditioned Krylov methods.}}
%
{
Radiative transfer calculations %simulations
for polarized radiation are performed in semi-empirical 1D solar atmospheres, out of local thermodynamic equilibrium (LTE).
%To obtain the population of the lower level,
A two-step solution strategy is applied: the non-LTE RT problem is first solved considering a multilevel atom and neglecting polarization phenomena.
The same problem is then solved including polarization and magnetic fields, considering a two-level atom and keeping fixed the population of the lower 
level calculated at the previous step.
%The ensuing linear problem 
The problem of step two is linear and it is solved with a preconditioned \mbox{FGMRES} iterative method.
%considering a two-level atom in the presence of a magnetic field and including polarization.
The emergent fractional linear polarization signals calculated %with CRD PRD--AA, 
under the CRD and PRD--AA approximations are analyzed and compared to those obtained by modeling PRD effects in their general angle-dependent (AD) 
formulation.
%and PRD--AD calculations are compared and analyzed.}
}
%
% {\textcolor{orange}{
% We find that the forward-scattering Hanle effect signals resulting from 
% angle-dependent PRD calculations can be one order of magnitude larger than 
% those obtained either assuming CRD or considering PRD effects under the 
% angle-average approximation.}}
{
With respect to the PRD--AD case, the
CRD and PRD--AA calculations significantly underestimate the amplitude of the line-center polarization signals
produced by the forward-scattering Hanle effect.
}
{
The results of this work suggest that a PRD--AD modeling is required in order to develop reliable diagnostic techniques exploiting the 
forward-scattering polarization signals observed in the Ca~{\sc i} 4227\,{\AA} line.
These results need to be confirmed by full 3D calculations including non-magnetic symmetry-breaking effects. 
}

\keywords{Magnetic fields -- Polarization -- Radiative transfer -- Scattering -- Sun: chromosphere}

\maketitle

\section{Introduction}

When observed in quiet or moderately active regions of the solar disk, many lines of the solar spectrum, from the near infrared to the far ultraviolet, show 
measurable linear polarization signals, produced by the scattering of anisotropic radiation 
\citep[e.g.,][]{stenflo1997sss,gandorfer2000,gandorfer2002,gandorfer2005,kano2017,rachmeler2022}.
These scattering polarization signals are receiving increasing attention by the scientific community because they encode %a wealth of 
precious information on the thermodynamic and magnetic properties of the solar atmosphere 
%, the reason their observation and modeling is receiving great interest in the scientific community 
\citep[e.g.,][and references therein]{jtb2022ar}.

When observed with low spatio-temporal resolution, scattering polarization signals are generally strongest
at the edge of the solar disk (limb) and decrease towards the disk center, where they typically vanish.
Consistently with this observational evidence, 
%theoretical calculations show that the amplitude of the scattering polarization signals
%{produced by atoms pumped by a radiation field with axial symmetry around the vertical}
it can be theoretically shown that in a setting with cylindrical symmetry around the vertical, 
the amplitude of scattering polarization scales as $(1 - \mu^2)$, with $\mu$
the cosine of the inclination of the line-of-sight with respect to the vertical \citep[e.g.,][]{LL04}.\footnote{We note that $\mu$ also corresponds to
the cosine of the heliocentric angle of the observed region.}
%However, a lack of axial symmetry in the %radiation field 
%problem can produce non-negligible scattering polarization signals
On the other hand, if the problem is not axially symmetric, %axial symmetry of the problem is broken,} 
non-negligible scattering polarization signals can also be obtained %are also obtained} 
at $\mu=1$, and thus also %in particular
%, in the forward-scattering geometry as when observing 
at the solar disk center, in a forward-scattering geometry \citep{jtb2001hanle}.
The first detection of %solar 
forward-scattering polarization on the Sun was achieved by \citet{jtb2002nat}
while observing a quiescent filament in the He~{\sc i} 10830\,{\AA} triplet. 
Shortly afterward, while pointing an active region, 
\citet{stenflo2003spw3} detected a forward-scattering signal %in an active region 
in the chromospheric Ca~{\sc i} line at 4227\,{\AA}.
Other remarkable disk-center observations in this line were %subsequently
carried out by \citet{bianda2011fshe} 
using the Zurich Imaging Polarimeter \citep[ZIMPOL-III,][]{ramelli2010zimpol}.

% A lack of axial symmetry in the radiation field
%{A lack of axial symmetry in the properties of the solar atmosphere, yielding a non-axially symmetric radiation field} 
A lack of axial symmetry in the solar atmosphere, 
capable of producing forward-scattering polarization, can be due to
(i) horizontal inhomogeneities in the density and temperature of the solar plasma \citep{manso2011},
(ii) spatial gradients of the plasma bulk velocity \citep[e.g.,][]{stepan2016,delpino2018}, and 
(iii) a deterministic magnetic field inclined with respect to the vertical \citep{jtb2001hanle,jtb2002nat}.
In the first two scenarios, the signals are produced by a lack of axial symmetry in the radiation field that illuminates the atoms; %, 
in the third one, generally referred to as forward-scattering Hanle effect, they arise directly from the magnetic field. 
% \eab{[\textit{Alternatively, simply remove everything after ``illuminates the atoms''. The magnetic origin of signals in case (iii) should be clear from the previous sentence}.]} 
%The latter scenario is generally referred to as the forward-scattering Hanle effect.
In general, these symmetry-breaking effects operate simultaneously, and the individual contributions can hardly be isolated.
Interestingly, radiative transfer (RT) calculations carried out in state-of-the-art 3D models of the solar atmosphere,
in the limit of complete frequency redistribution (CRD), show that the forward-scattering signals produced 
by the symmetry breaking effects (i) and (ii) are reduced by the presence of magnetic fields \citep{stepan2016,delpino2018,jaume2021CaI}.
Moreover, it can be argued that the inherent averaging happening in observations with low spatio-temporal resolution significantly
conceals the impact of
mechanisms (i) and (ii), and indeed it is currently rather challenging to detect forward-scattering polarization signals in very quiet solar regions.
As a matter of fact, the aforementioned observations in the Ca~{\sc i} 4227\,{\AA} line were all performed in relatively strongly magnetized regions (i.e., with 
noticeable $V/I$ signals), suggesting that the forward-scattering Hanle effect might be the main mechanism responsible for their generation.
To routinely detect forward-scattering polarization signals in quiet solar regions is one of the challenges for
%\textcolor{magenta}{By combining a high polarimetric sensitivity with high spatio-temporal resolutions,}
the new generation of large-aperture solar telescopes, such as DKIST \citep{rimmele20} and the future EST \citep{quintero22}.
%%combine a high polarimetric sensitivity with high spatio-temporal resolutions.
%%Such facilities 
%should be able to routinely detected forward-scattering polarization signals 
%\textcolor{magenta}{\st{produced by all the above-mentioned symmetry-breaking effects}} also in quiet solar regions.

In order to interpret the available observational data,
various theoretical works have already been carried out on the modeling of forward-scattering polarization signals %, especially for the 
in the chromospheric Ca~{\sc i} 4227\,{\AA} line.
%Considering 1D semi-empirical solar atmospheric models, \citet{anusha2011fshe} inferred information on low-chromospheric magnetic fields by
%modeling the observations of \citet{bianda2011fshe}.
%Although exploiting the \emph{angle-averaged} (AA) simplifying approximation \citep[see][Sect.~4.3]{bommier1997b}, this work 
\citet{anusha2011fshe} modeled the observations of \citet{bianda2011fshe} in 1D semi-empirical solar atmospheric models, 
taking partial frequency redistribution (PRD) effects into account under the angle-average (AA) approximation 
\citep[see][Sect.~4.3]{bommier1997b}.
Their work allowed inferring information on the magnetic fields of the low chromosphere, and showed 
%and highlighted the importance of including partial frequency redistribution (PRD) effects, showing 
that various observed features cannot be reproduced if the CRD limit
% of complete frequency redistribution (CRD) 
is considered.
Subsequently, \citet{carlin2016fshe,carlin2017fshe} modeled the Ca~{\sc i} 4227 forward-scattering polarization signals in 3D models of the solar atmosphere,
%exploiting 
in the limit of CRD. %and neglecting
Their calculations neglected both the effects of horizontal RT (i.e., they considered the so-called 1.5D approximation) 
and the horizontal component of the model's bulk velocity.
This work highlighted the key role played by the dynamics and temporal evolution of the chromospheric plasma during the integration time of the observations.
In the investigations mentioned so far, the symmetry-breaking mechanisms (i) and (ii) were not taken into account.
By contrast, \citet{jaume2021CaI} investigated the polarization of the Ca~{\sc i} 4227\,{\AA} line performing full 3D calculations in realistic models of the 
solar atmosphere
with the radiative transfer code PORTA
\citep{stepan2013} under the assumption of CRD.
Accounting for all symmetry-breaking effects,
their study showed that the spatial gradients in the horizontal component of the plasma bulk velocity can produce 
conspicuous forward-scattering polarization signals, comparatively larger than those produced by horizontal inhomogeneities in the solar plasma, and that 
these signals are reduced by the presence of a magnetic field.

So far, all the theoretical investigations on the forward-scattering polarization of the Ca~{\sc i} 4227\,{\AA} line have been
carried out either in the limit of CRD or considering PRD effects under the AA simplifying approximation.
This is ultimately due to the formidable computational complexity of modeling scattering polarization while accounting for PRD phenomena in their most general
angle-dependent (AD) formulation.
Performing RT calculations in realistic atmospheric models, taking AD PRD effects into account, is today feasible
\citep[e.g.,][]{delpino20,benedusi2023}, and a series of 1D applications to Ca~{\sc i} 4227 
have been presented \citep[e.g.,][]{janett2021a,riva_2023_RIIIEx,guerreiro24}.
However, none of these works specifically investigated the forward-scattering Hanle effect.
The present study focuses on this effect and aims at assessing the suitability of the CRD and PRD--AA approximations 
for its modeling in the Ca~{\sc i} 4227\,{\AA} line.

The article is organized as follows: Section~\ref{sec:problem} exposes the considered RT problem for polarized radiation and the adopted solution strategy.
In Sect.~\ref{sec:results}, we report and analyze the emergent fractional linear polarization of the Ca~{\sc i} 4227 line, comparing CRD, PRD--AA, and 
PRD--AD calculations.
Section~\ref{sec:discussion} discusses the main results and their implications, while Sect.~\ref{sec:conclusions} provides remarks and conclusions.

\section{Problem formulation and solution strategy}
\label{sec:problem}

The Ca~{\sc i} %spectral 
line at 4227\,{\AA} is produced by the transition between the ground level of neutral calcium, $4s^2 \, ^1\rm{S}_0$, and the excited level
$4s4p \, ^1\rm{P}^{\rm{o}}_1$. 
In quiet regions close to the solar limb, this line shows a large scattering polarization signal, characterized by broad lobes in the
wings and a sharper peak in the core \citep[e.g.,][]{gandorfer2002}.
Both the line-core peak and the wing lobes are sensitive to the magnetic field, the former through the Hanle effect 
\citep[which only operates in the line-core region, e.g.,][]{LL04} and the latter through magneto-optical effects \citep{alsina2018CaI}.
In low-resolution observations, the forward-scattering polarization signals detected close to the disk center consist 
instead in a single peak in the line-core region \citep[e.g.,][]{bianda2011fshe}.
A correct modeling of the scattering polarization profiles of this line, and in particular of the wing lobes, 
requires taking PRD effects into account \citep[e.g.,][]{faurobert88}.
Nonetheless, the limit of CRD can be used to approximately model the line-core peak, in both limb and forward-scattering geometries 
\citep[e.g.,][]{sampoorna2010,jaume2021CaI}.
%As a matter of fact, in low-resolution observations, the forward-scattering polarization signals detected close to the disk center consist 
%in a single peak in the line-core region \citep[e.g.,][]{bianda2011fshe}.

%The scattering polarization signal of this line can be suitably modeled considering a two-level atom with an unpolarized and infinitely sharp lower level
%This allows us to account for PRD effects by applying the redistribution matrix for this atomic model as derived by \citet{bommier1997a,bommier1997b}.
The scattering polarization signal of the Ca~{\sc i} 4227\,{\AA} line can be suitably modeled considering a two-level atom.
Indeed, the line is not part of a multiplet (the upper and lower levels have spin zero) and, since the upper level is not connected to lower-energy 
levels other than the ground level, its population is mainly determined by this line itself (assuming that the ionization fraction is known).
%We also note that the lower level, having $J=0$, cannot carry atomic polarization and that, being the long-lived ground level of neutral calcium, it can 
%be treated as infintely sharp.
In this work, we account for PRD effects using the redistribution matrix for a two-level atom with unpolarized and infinitely-sharp lower level as derived by
\citet{bommier1997a,bommier1997b}.
The long-lived ground level of neutral calcium can indeed be treated as infinitely sharp and, having total angular momentum $J=0$, it cannot 
carry atomic polarization.
%This
The redistribution matrix is given by the sum of two terms: one that describes scattering processes that are coherent in the atomic reference frame ($R^{\scriptscriptstyle \mathrm{II}}$), and the other describing scattering processes that are totally incoherent in the same reference frame ($R^{\scriptscriptstyle \mathrm{III}}$). 
In the observer's reference frame, we consider the exact AD expression of $R^{\scriptscriptstyle \mathrm{II}}$, while we make the assumption of totally 
incoherent scattering for $R^{\scriptscriptstyle \mathrm{III}}$ \citep[e.g.,][]{bommier1997b,sampoorna2017,riva_2023_RIIIEx}.
The CRD modeling is instead performed by applying the theoretical approach described in \citet{LL04}.

In this work, we only focus on the forward-scattering Hanle effect, neglecting the two other mechanisms (i) and (ii).
We thus consider the static 1D semi-empirical atmospheric \textit{model C} of \citet[][hereafter FAL-C]{fontenla1993}, and we analyze the impact of 
inclined deterministic magnetic fields.
%To investigate the sole impact of a deterministic magnetic field on forward-scattering polarization signals from the two other mechanisms (i) and (ii), here we consider the
%static 1D semi-empirical atmospheric \textit{model C} of \citet[][hereafter FAL-C]{fontenla1993}.
%% \jiri{[Here I would not use the word ``disentangle'' because I understand it as a procedure to be used in the full 3D atmosphere with all the contributing mechanisms acting together. Maybe better to say something like ``To investigate the sole impact...''?]}
%We note that the considered approach and redistribution matrices are fully general and include the effect of both arbitrary magnetic and bulk velocity fields.
%However, to better focus on the forward-scattering Hanle effect alone, we consider a static setting, that is we neglect the impact of plasma bulk velocities.
%In the calculations, we analyze the impact of both horizontal and inclined magnetic fields.
%
The intensity and polarization profiles of the emergent radiation are calculated by solving the non-LTE RT problem for polarized radiation.
The adopted solution strategy requires two steps: first, we solve the non-LTE RT problem neglecting polarization phenomena.
This first step is carried out using the RH code \citep{uitenbroek2001}, and considering an atomic model for calcium composed of 25 levels 
(including five levels of Ca~{\sc ii} and the ground level of Ca~{\sc iii}).
% {\textst{The RH code output additionally provides the rates for elastic and inelastic collisions required in step 2.}}
Second, we solve
the non-LTE RT problem including polarization, but keeping fixed the population of the lower level calculated in step 1, so that the problem is 
linear \citep[e.g.,][]{janett2021b}.
This is achieved
with a preconditioned \mbox{FGMRES} (Flexible Generalized Minimal RESidual) iterative method, as described in % by setting up a tolerance of $10^{-9}$}
\citet[][]{benedusi2021,benedusi2022a}
and~\citet{janett2024}.
All the physical and numerical parameters are the same as in \citet{guerreiro24}.
% {\textst{The profiles and off-grid solutions (used in this work) were computed by performing a formal solution along a given LOS by using the converged solutions as initial guesses.}}

\section{Results}
\label{sec:results}

This section presents the results of our calculations
of the linear scattering polarization profiles
of the Ca~{\sc i} line at 4227~{\AA}
both in the absence and in the presence
of magnetic fields.
The problem is formulated in a right-handed Cartesian coordinate system with the $z$-axis directed along 
the vertical, and the $x$-axis directed so that the line-of-sight (LOS) lies in the $x-z$ plane ($x>0$, $z>0$ quadrant).
The LOS is thus fully specified by the cosine of its inclination $\theta$ with respect to the vertical, that is $\mu = \cos(\theta) \in [0,1]$.
%The inclination $\theta$ also corresponds to the heliocentric angle of the observed point.
The reference direction for positive Stokes $Q$ is taken parallel to the $y$-axis.
The magnetic field vector is specified by the intensity $B$, the inclination 
$\theta_B$ with respect to the vertical, and the azimuth $\chi_B$ measured on 
the $x-y$ plane, counter-clockwise (for an observer at $z>0$) from the $x$-axis.
In Sect.~\ref{sec:horizontalB}, we analyze the case of horizontal (i.e., $\theta_B=\pi/2$) magnetic fields,
while the case of non-horizontal inclined magnetic fields (i.e., $\theta_B=\pi/4$)
is considered in Sect.~\ref{sec:inclinedB}.

\begin{figure}[!ht]
	\centering
    \includegraphics[width=0.5\textwidth]{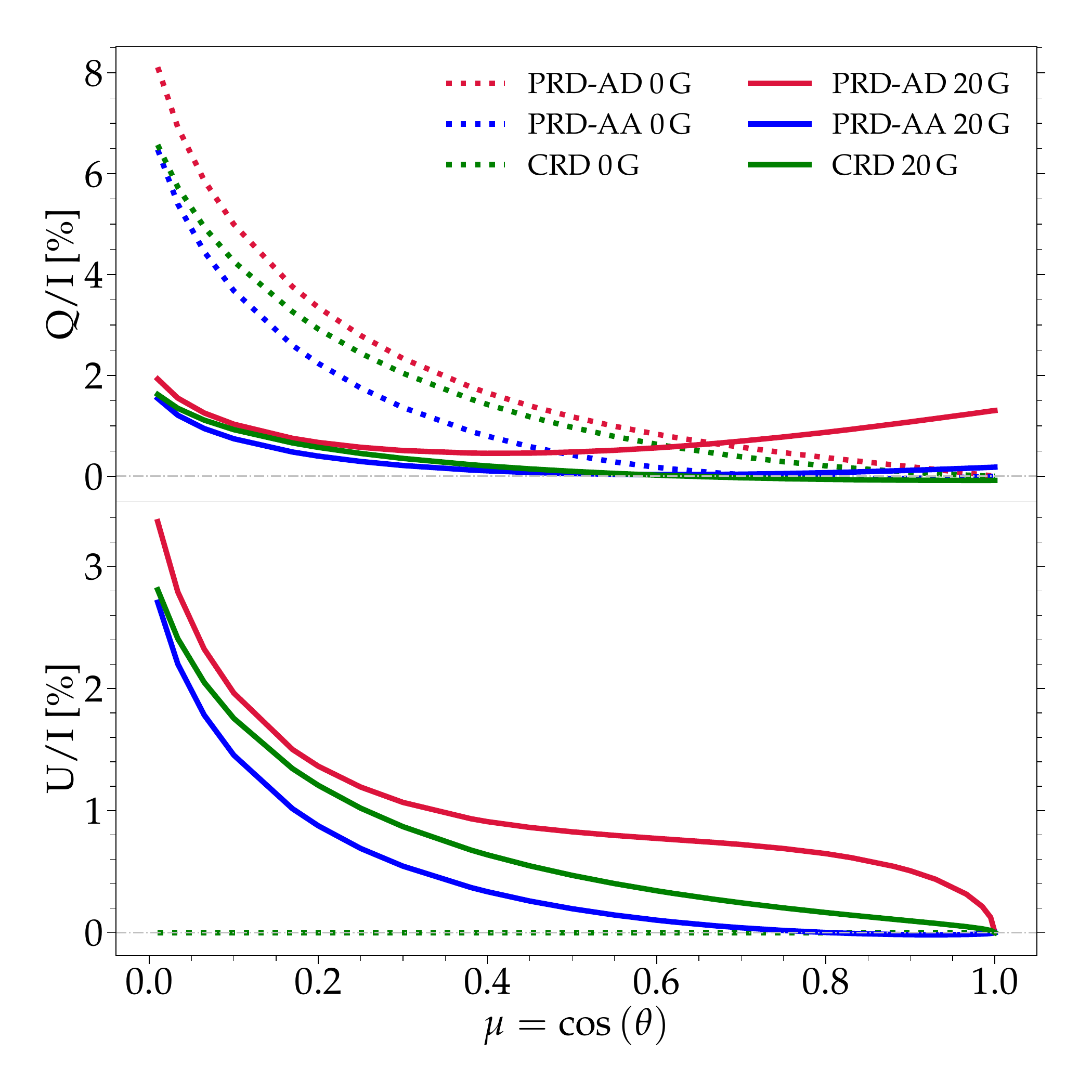}
%    \vspace*{-0.8cm}
    \caption{Center-to-limb variation of the Ca~{\sc i} 4227 line-center
    fractional linear polarization $Q/I$ (\emph{upper panel}) and $U/I$ (\emph{lower panel}), 
    obtained from CRD, PRD--AA, and PRD--AD calculations in the FAL-C atmospheric model, 
    both in the absence (\emph{dotted curves}) and in the presence (\emph{solid curves}) 
    of a magnetic field.
    The dotted curves for $U/I$ are equal to zero.
    In the magnetic case, a height-independent horizontal ($\theta_B=\pi/2$, $\chi_B=0$) magnetic field of 20\,G is considered. The reference direction for positive Stokes $Q$ is parallel to the $y$-axis of the considered reference system, that is perpendicular to the magnetic field.}
\label{fig:clv}
% \end{figure}
% %
% \begin{figure}
	% \centericorng
    \includegraphics[width=0.5\textwidth]{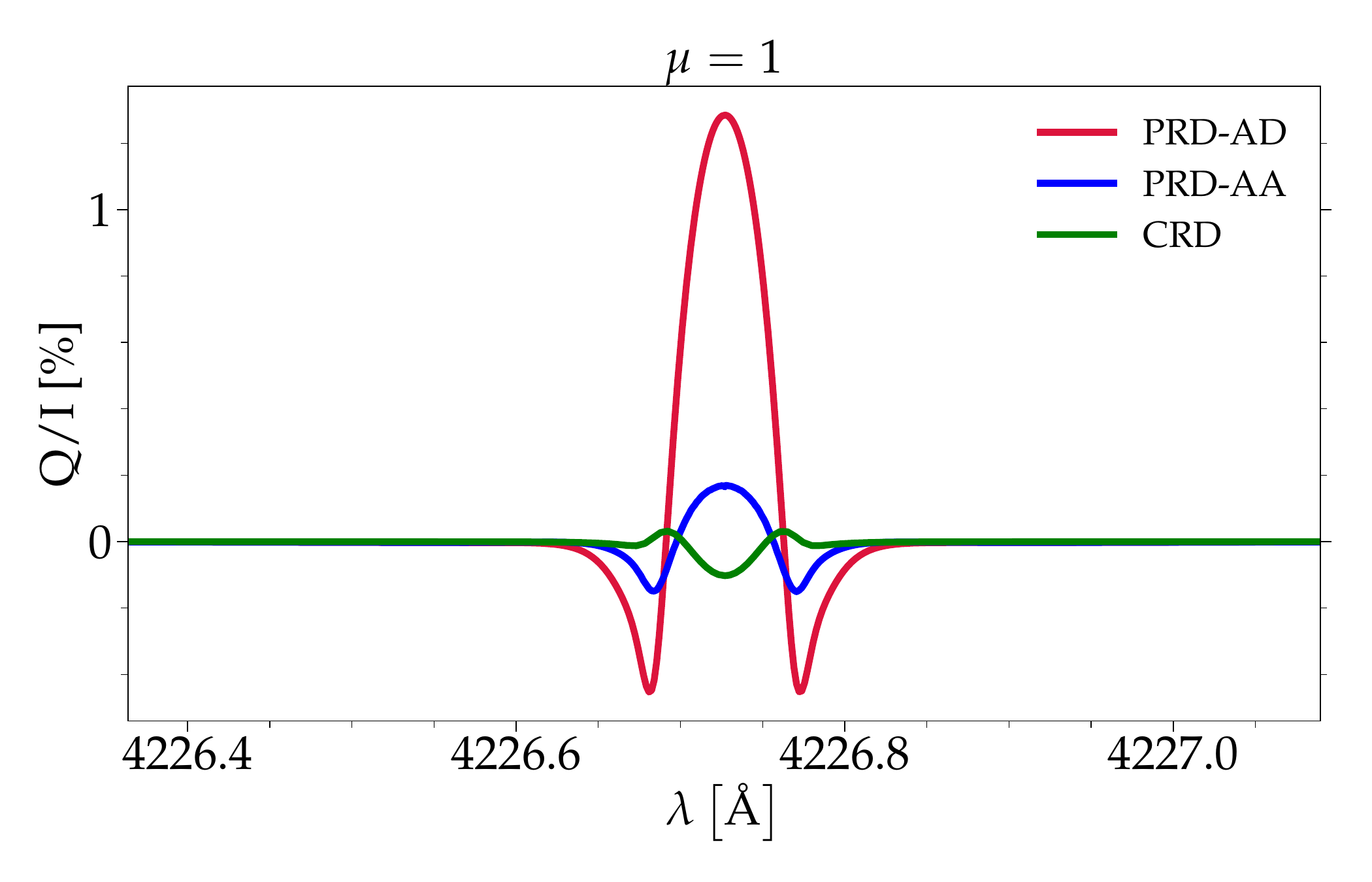}
    \caption{Emergent $Q/I$ profiles of Ca~{\sc i} 4227 %line fractional linear polarization $Q/I$ 
    at $\mu=1$, obtained from CRD, PRD--AA, and PRD--AD calculations in the FAL-C atmospheric model, 
    in the presence of a height-independent horizontal ($\theta_B=\pi/2$, $\chi_B=0$) magnetic field of 20\,G.
    The reference direction for positive Stokes $Q$ is parallel to the $y$-axis of the considered reference system, 
    that is, perpendicular to the magnetic field.
    In this geometry, the $U/I$ profile vanishes, and it is thus omitted.}
    \label{fig:profiles}
\end{figure}
\begin{figure*}[ht]
	\centering
    \includegraphics[width=0.9\textwidth]{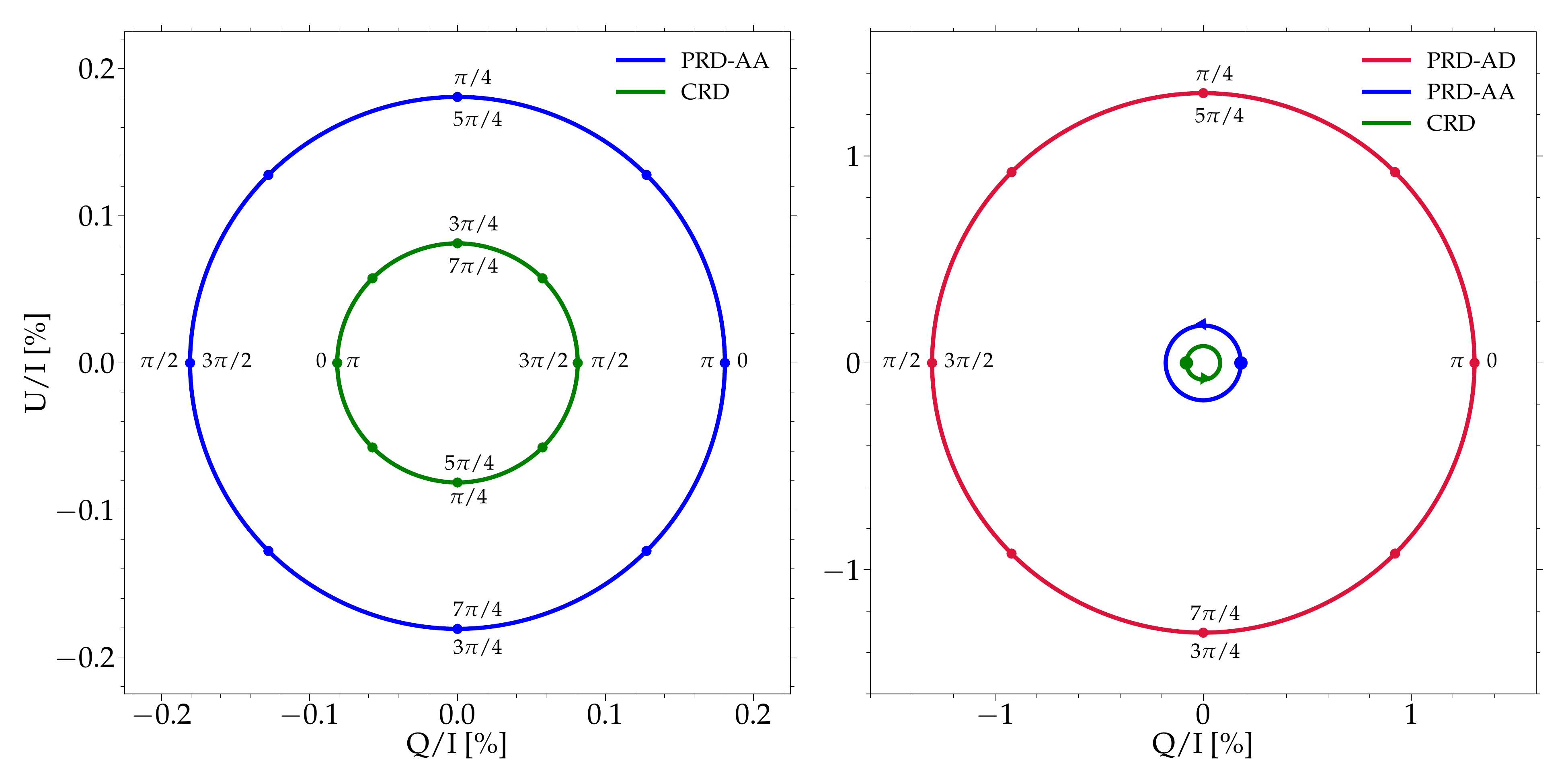}
    \caption{Polarization diagrams of the Ca~{\sc i} 4227 line-center emergent radiation at $\mu=1$,
    obtained from CRD, PRD--AA, and PRD--AD calculations in the FAL-C atmospheric model, 
    in the presence of a height-independent horizontal ($\theta_B = \pi/2$) magnetic field of 20\,G.
    The circular markers indicate the effective calculations, %which considered 
    carried out for the azimuths $\chi_B = n \, \pi/8$ (with $n=0,...,15$).
    \emph{Left panel}: comparison between CRD and PRD--AA calculations. 
    \emph{Right panel}: comparison between CRD, PRD--AA, and PRD--AD calculations 
    (in the PRD--AA and CRD diagrams the circular marker indicates a magnetic field direction 
    with azimuth $\chi_B = 0$, whereas the arrow refers to $\chi_B=\pi/4$).
    The reference direction for positive Stokes $Q$ is parallel to 
    the $y$-axis of the considered reference system.}
	\label{fig:hanle}
\end{figure*}

\subsection{Horizontal magnetic fields}
\label{sec:horizontalB}

Figure~\ref{fig:clv} shows the center-to-limb variation of the amplitude
of the line-center scattering polarization $Q/I$ (upper panel) and 
$U/I$ (lower panel) signals
of the Ca~{\sc i} 4227\,{\AA} line, 
resulting from CRD, PRD--AA, and PRD--AD calculations.
As expected, in the absence of magnetic fields (dotted curves), the amplitude of the $Q/I$ peak monotonically decreases while 
moving from the limb towards the disk center, where it vanishes. 
The $U/I$ signal is always zero, consistently with our choice of the reference direction for positive Stokes $Q$.
For any limb distance, the largest signals are obtained in the PRD--AD setting.

The presence of a height-independent horizontal ($\theta_B=\pi/2$, $\chi_B=0$) magnetic field of 20\,G produces a significant Hanle 
rotation at the limb.
This leads to a depolarization of the line-center $Q/I$ signal 
while giving rise to an appreciable $U/I$ %Hanle 
signal (see solid curves).
At the solar disk center, the forward-scattering Hanle effect yields a non-zero $Q/I$ signal, 
while $U/I$ is zero, consistently with the geometry of the problem.
% and the chosen reference direction for positive Stokes $Q$.  
The CRD, PRD--AA, and PRD--AD calculations show significant differences for all $\mu$, 
in both $Q/I$ and $U/I$, 
with the PRD--AD results always showing the largest signal. 
A very interesting finding is that the PRD--AD results for $Q/I$ drift away from the others while approaching the disk center.
At $\mu=1$, the PRD--AD signal is well above 1\%, that is one order of magnitude larger than in the PRD--AA and CRD cases.
Moving from the limb to the disk center, the $Q/I$ signal obtained from PRD--AA and PRD--AD calculations first decreases, reaches a minimum, and then increases, 
always remaining positive.
On the contrary, in the CRD case, it shows a sign reversal at around $\mu \approx 0.6$ and then remains negative until $\mu=1$, with amplitudes always 
below 0.1\% (see also Fig.~\ref{fig:profiles}).
We note that the
% sensibly lower amplitudes showed by PRD--AA calculations 
differences between PRD--AA and PRD--AD calculations in Fig.~\ref{fig:clv}
are magnified by the artificial depolarization that is
% \MyNewEdit{We note that the significantly lower amplitudes shown in PRD--AA calculations are due to the non-physical artifacts  }
introduced by the AA approximation in the line-core
of $Q/I$ and $U/I$ for $\mu \ne 1$, both in the absence and presence of magnetic fields
\citep[see][]{janett2021a}.
To better visualize the impact of the different scattering modelings on the forward-scattering Hanle effect signal, Fig.~\ref{fig:profiles}
shows the emergent $Q/I$ profiles at $\mu=1$.
This clearly highlights the significantly stronger polarization signals resulting from PRD--AD calculations.
% We note that the small negative peaks in the near wings are typical features that appear when solving the RT problem in an optically thick medium with a photon destruction probability that is not too low \citep[e.g.,][]{jtb1999,LL04}

\begin{figure}[ht!]
    \centering
    \includegraphics[width=0.5\textwidth]{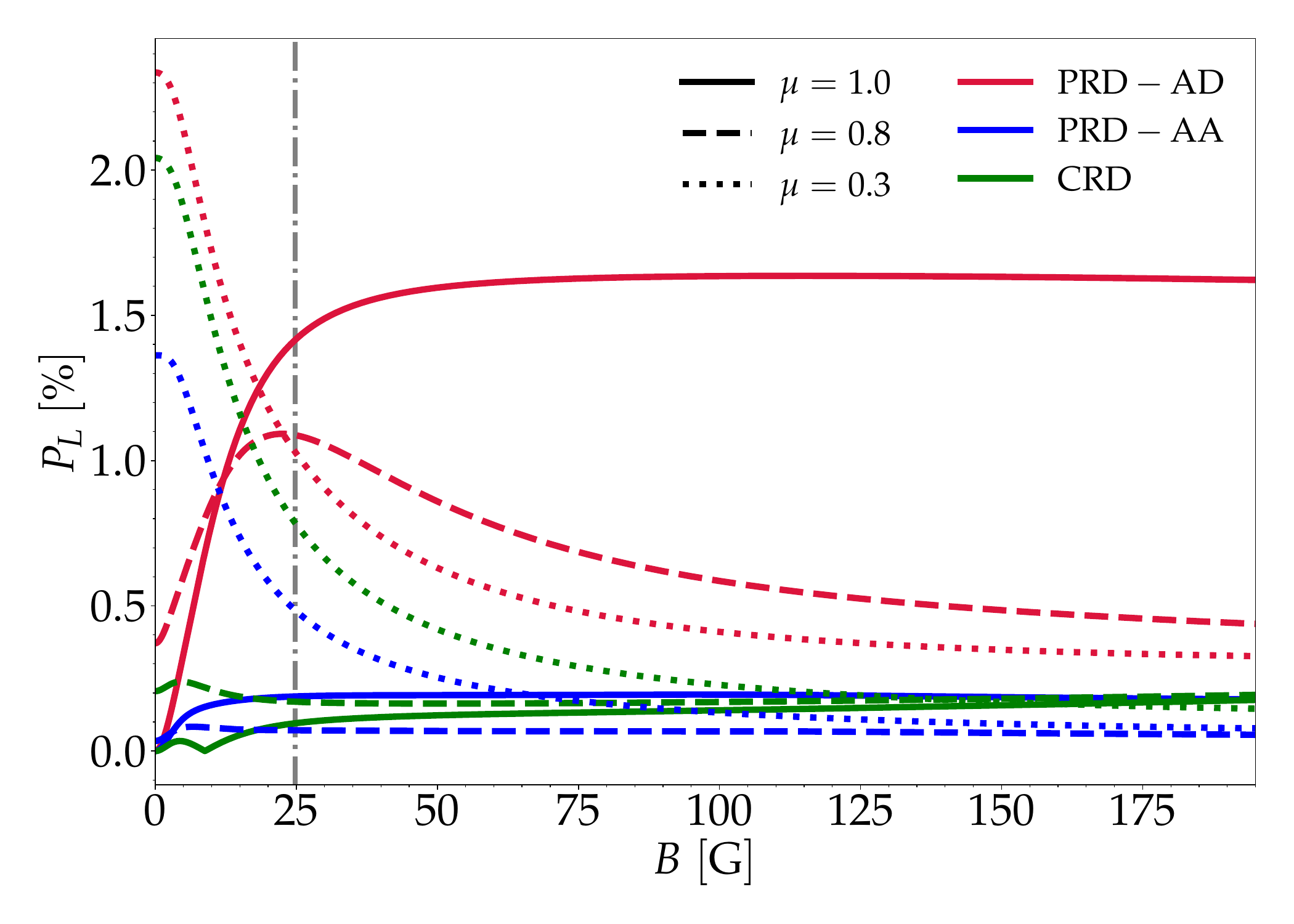}
    \caption{Line-center %frequency 
    linear polarization degree %~\eqref{eq:PL} 
    as a function of the strength of a horizontal
    ($\theta_B=\pi/2$, $\chi_B=0$) magnetic field, %for different LOSs, 
    obtained with PRD--AD (\emph{red}), PRD--AA (\emph{blue}) and CRD (\emph{green}) calculations, for different LOSs.
    % The solid and dashed  lines are results for LOSs $\mu \in \{ 0.8, \, 1 \}$ (FSHE regime), while the dotted are for $\mu = 0.3$. 
    The vertical dash-dotted line indicates the Hanle critical field of Ca~{\sc i} 4227, that is $B_H=25$\,G.}
    \label{fig:PL}
\end{figure}

Figure~\ref{fig:hanle} compares the polarization diagrams for the line-center radiation emitted at $\mu=1$, obtained with PRD--AD, PRD--AA and CRD calculations, considering
height-independent horizontal (i.e., $\theta_B=\pi/2$) magnetic fields of 20\,G with different azimuths $\chi_B$.
These diagrams further highlight
the large polarization amplitudes of the 
forward-scattering Hanle effect signals obtained in the PRD--AD setting.
We find differences of approximately one order of magnitude (in both $Q/I$ and $U/I$) when comparing PRD--AD with PRD--AA and CRD results.
Much smaller, yet relevant, differences are instead found between the CRD and PRD--AA calculations.
Consistently with the geometry of the problem, $Q/I$ and $U/I$ vanish for $\chi_B=\pi/4 + n \, \pi/2$ %,3\pi/4,5\pi/4,7\pi/4$,
and $\chi_B=n \, \pi/2$ %0,\pi/2,\pi,3\pi/2$.
($n=0,...,3$), respectively.
It can be noticed that the evolution of the amplitudes of $Q/I$ and $U/I$ shows a counter-clockwise 
rotation in all cases.
Interestingly, the line-center signal provided by CRD calculations
always has the opposite sign with respect to the one
provided by PRD--AA and PRD--AD calculations.

% Finally, 
Figure~\ref{fig:PL} shows the line-center %frequency
linear polarization degree
\begin{equation}
    P_L   = \frac{ \sqrt{ Q^2  +  U^2 }}{I},
    \label{eq:PL}
\end{equation}
for different LOSs and magnetic field strengths 
of a horizontal magnetic field ($\theta_B=\pi/2$, $\chi_B=0$),
obtained with PRD--AD, PRD--AA, and CRD calculations.
At the solar disk center ($\mu=1$, solid lines), where the polarization is fully produced by the forward-scattering Hanle effect, 
$P_L$ is already appreciable for magnetic fields of a few gauss only.
Its value quickly grows %by further increasing the magnetic field strength
as the magnetic field strength increases further, and finally stabilizes above the Hanle critical field.
% \textcolor{red}{*** Simone: Anche questo $\mu=1$ sembra un regime di saturazione ***}
The increase of $P_L$ is much steeper and significant in the PRD--AD case than in the PRD--AA and CRD ones.
Notably, the increase is not monotonic in the CRD case. 
As expected, for a near-limb LOS ($\mu=0.3$, dotted lines), the Hanle effect produces a monotonic decrease of $P_L$ 
with the magnetic field strength, until reaching a saturation regime.
For an intermediate LOS ($\mu=0.8$, dashed lines), the value of $P_L$ first increases with the field strength, it reaches a maximum,
and it finally decreases till the saturation regime.
In the PRD--AD case, 
% \MyNewEdit{in the FSHE geometry ($\mu \in \{ 0.8 , 1 \}$) },
the initial increase is much steeper and significant, and
% \MyNewEdit{a change of behaviour begins around the Hanle critical field} 
the maximum is reached for stronger fields (around the 
Hanle critical field) than in the other cases.
The PRD--AD calculations show significantly larger values of $P_L$ for all LOS and for any magnetic field strength.

%
% \textcolor{red}{*** Mettere da qualche parte l'equazione di $B_H =  \frac{A_{u \ell}}{ 8.79 \cdot 10^{6}\; g_u }$ ****}
% {This figure shows that the FSHE is strictly correlated with the Hanle critical field (24.8\,G in the Ca~{\sc i} 4227 line, dashed vertical black line). 
% In fact in LOSs with direction $\mu \in \{ 0.8, \, 1 \}$ which are in FSHE regime, it is possible to observe that at $\mu=1$ (disk center) the value of $P_L$ grow quickly as function of the magnetic field strength up to the value of the critical field. While for larger values it stabilizes to a near constant value.
% For $\mu=0.8$ (near the disk center and in the FSHE regime) it is possible to observe that the maximal $P_L$ is reached for a value that is closed to the critical field, and, after that it decrease due to the Hanle effect.
% While where the LOS is $\mu=0.3$ that is a regime dominated by the Hanle effect alone the value of $P_L$ is damped according to the Hanle effect up to the value of saturation (which, theoretically, corresponds to the $1/5$ of the value in the absence of a magnetic field).}

\subsection{Inclined magnetic fields}
\label{sec:inclinedB}

\begin{figure}[!t]
	\centering
    \includegraphics[width=0.5\textwidth]{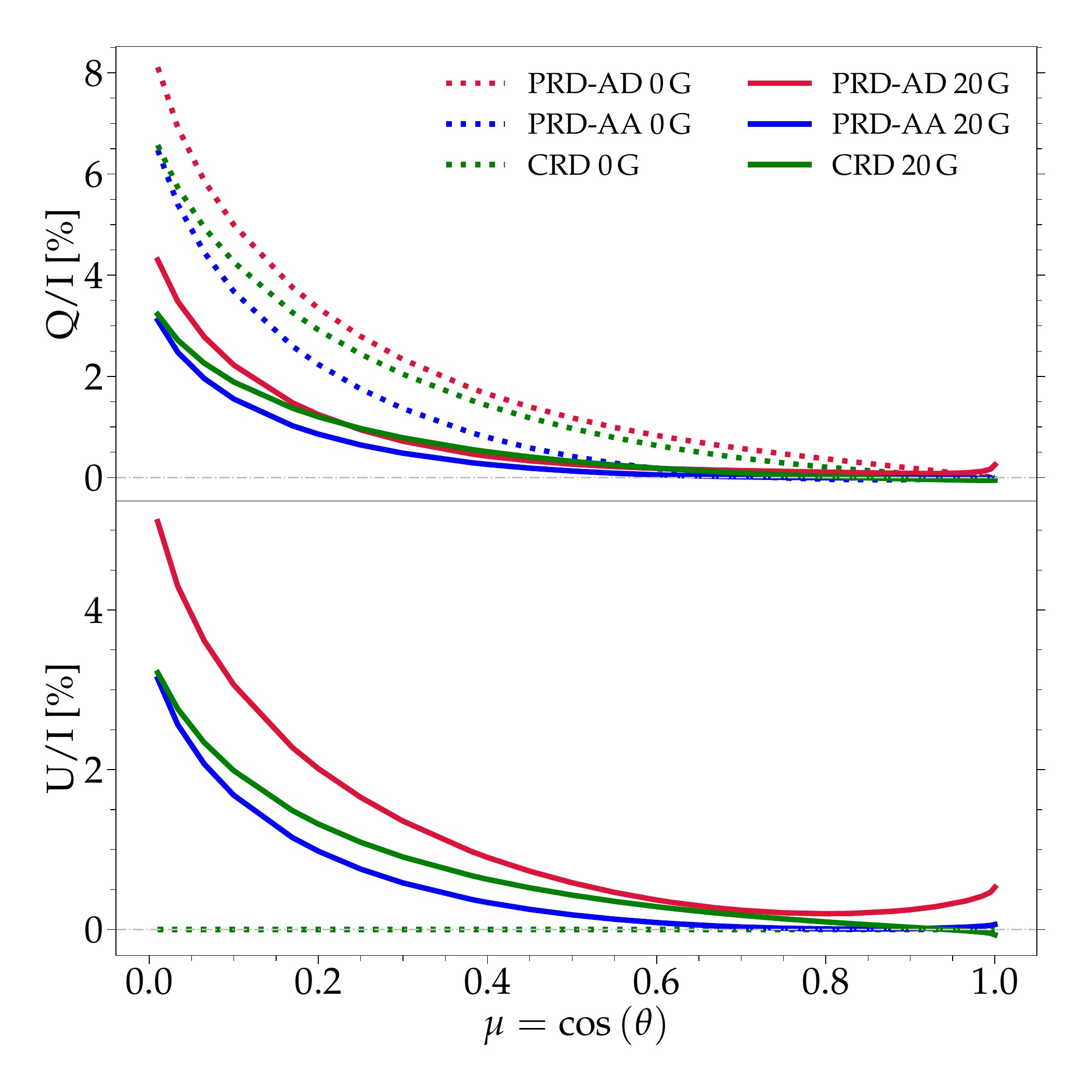}
%    \vspace*{-0.8cm}
    \caption{Same as Fig.~\ref{fig:clv}, but for a height-independent 
    inclined ($\theta_B=\pi/4$, $\chi_B=0$) magnetic field of 20\,G.}
    \label{fig:clv_pi4_pi4}
\end{figure}

\begin{figure}
	\centering
    \includegraphics[width=0.5\textwidth]{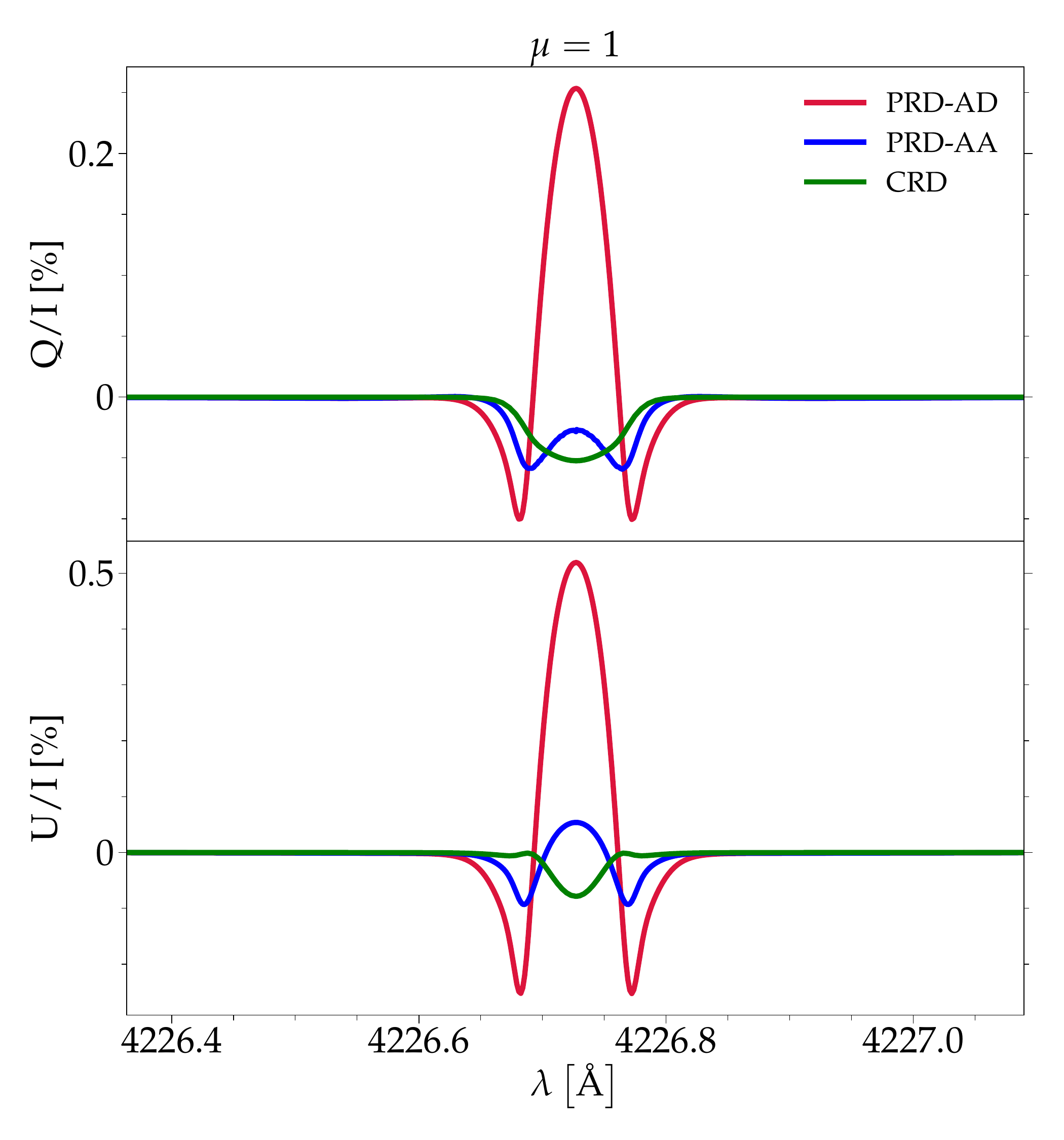}
	\caption{Same as Fig.~\ref{fig:profiles}, but for a height-independent 
    inclined ($\theta_B=\pi/4$, $\chi_B=0$) magnetic field of 20\,G.}
	\label{fig:profiles_BIncl}
\end{figure}

The results presented in the previous section are obtained considering horizontal magnetic 
fields. This geometry maximizes the breaking of the axial symmetry for a given value of $B$, and thus the 
amplitude of the polarization signals produced at the disk center via the 
forward-scattering Hanle effect.
On the other hand, the discrepancies between the emergent Stokes profiles 
resulting from PRD--AD, PRD--AA, and CRD calculations change significantly if 
magnetic fields with different inclinations with respect to the vertical are 
considered.
For this reason, we now analyze the case of a non-horizontal inclined magnetic field.
Figure~\ref{fig:clv_pi4_pi4} shows the center-to-limb variation of the amplitude
of the line-center scattering polarization $Q/I$ (upper panel) and 
$U/I$ (lower panel) signals of the Ca~{\sc i} 4227\,{\AA} line, 
resulting from PRD--AD, PRD--AA, and CRD calculations in the presence
of a height-independent inclined ($\theta_B = \pi / 4$, $\chi_B=0$) magnetic field of 20\,G.
This setting confirms the presence of relevant differences between 
the different scattering descriptions for all $\mu$,
and in particular for the forward-scattering geometry $\mu=1$.
To better visualize the impact of the
different scattering modelings on the
forward-scattering Hanle effect signal, Fig.~\ref{fig:profiles_BIncl}
shows the emergent $Q/I$ and $U/I$
profiles at $\mu=1$
for the same inclined magnetic field.
%Differently from Fig.~\ref{fig:profiles},
Unlike in the case of a horizontal magnetic field shown in
Fig.~\ref{fig:profiles}, 
this geometry
produces an appreciable $U/I$ signal.
% We note that 
The significantly stronger linear 
polarization signals resulting from PRD--AD calculations
are found both in $Q/I$ and $U/I$.

\begin{figure*}
	\centering
    \includegraphics[width=0.9\textwidth]{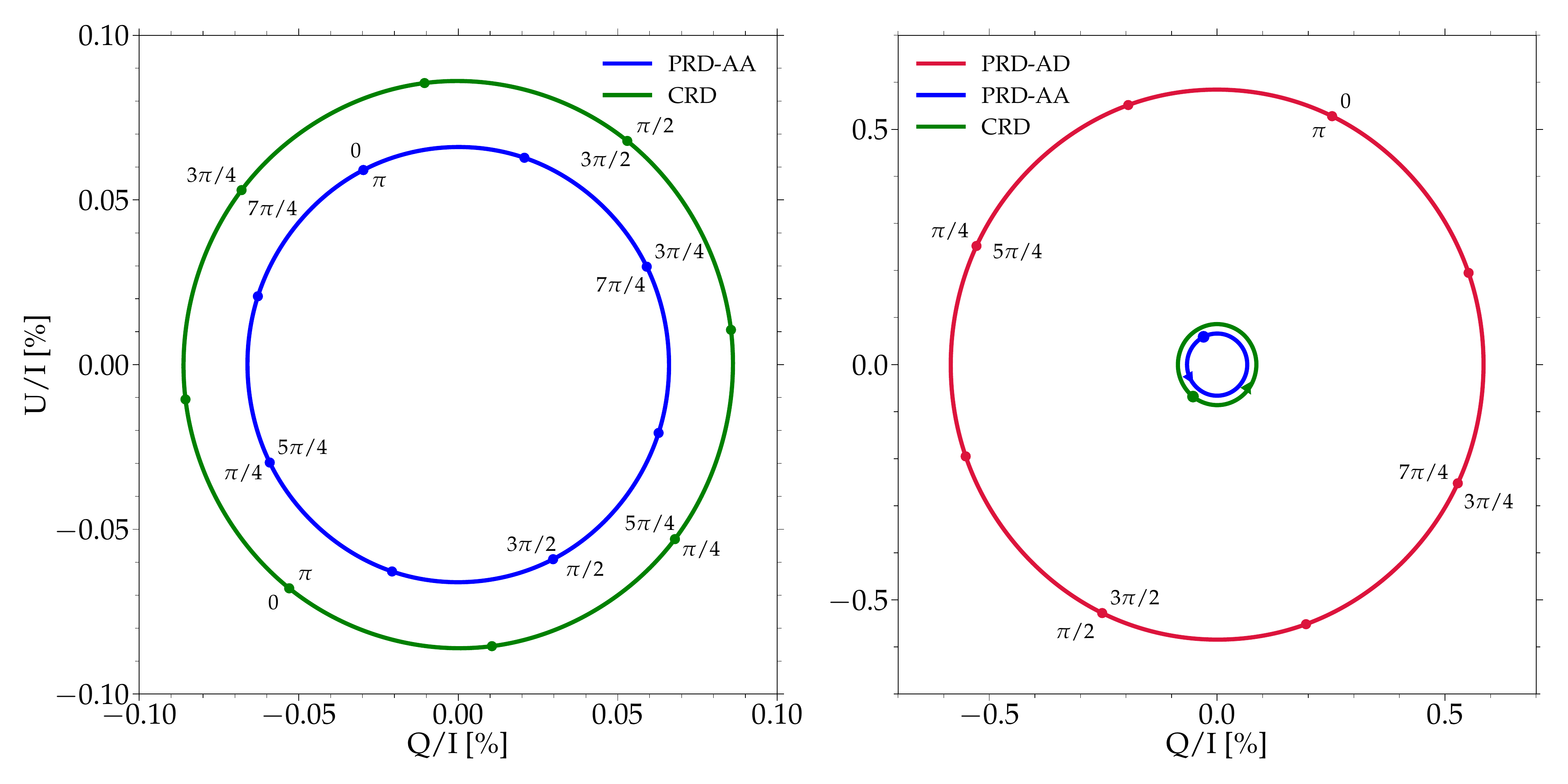}
    \caption{Same as Fig.~\ref{fig:hanle}, but for a height-independent 
    inclined ($\theta_B=\pi/4$) magnetic field of 20\,G.}
    \label{fig:hanleIncl}
\end{figure*}

\begin{figure*}
    \centering
    \includegraphics[width=0.9\textwidth]{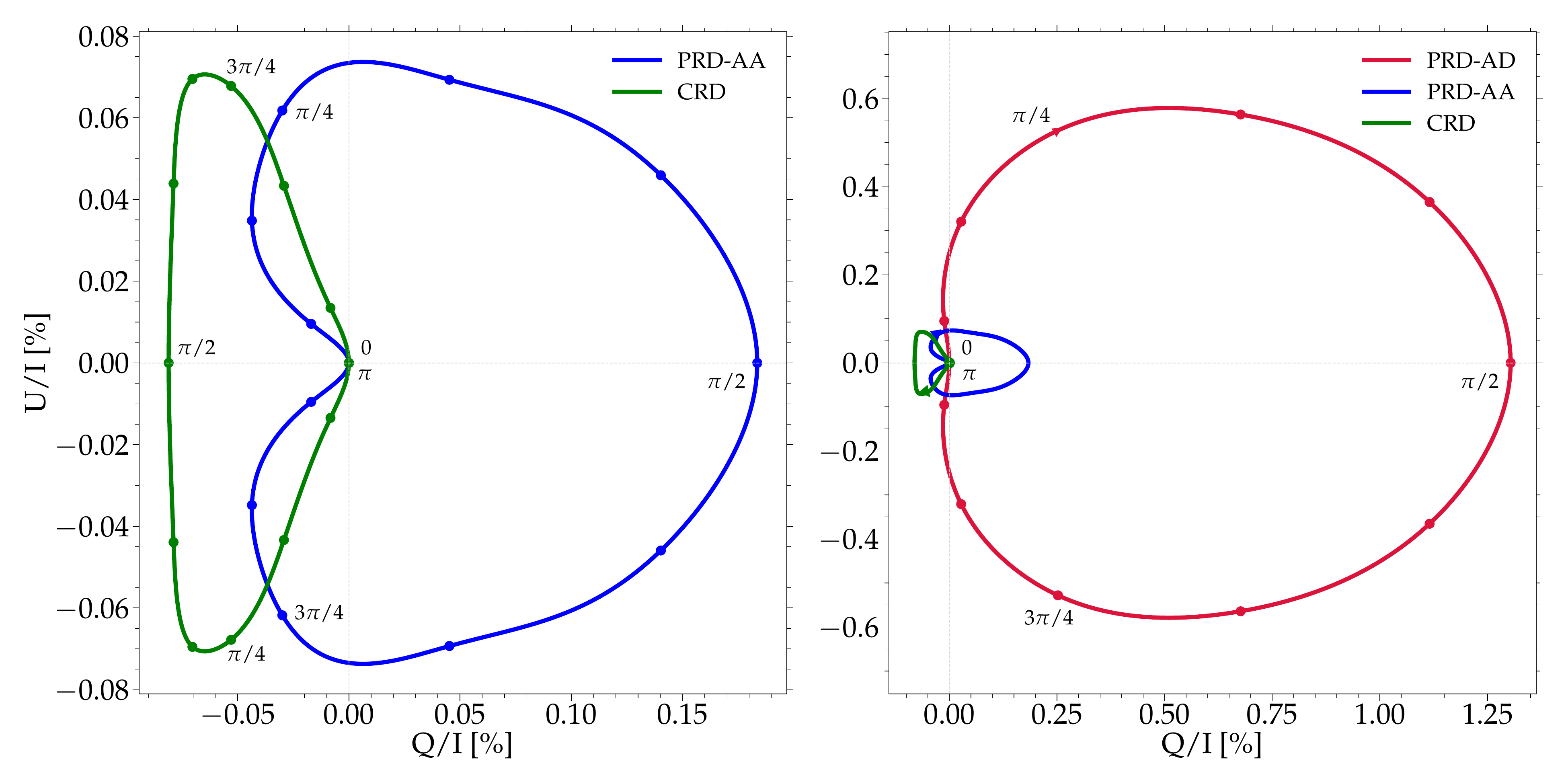}
    \caption{Polarization diagrams obtained considering a height-independent magnetic field 
    of 20\,G with a fixed azimuth $\chi_B = 0$, and varying its inclination $\theta_B \in [0,\pi]$. 
    The circular markers indicate the effective calculations, carried out for $\theta_B = n\, \pi/12$ with $n = 0, ..., 12$.
    The other parameters are the same as in Fig.~\ref{fig:hanle}.}
    \label{fig:hanleVar_theta}
\end{figure*}

Figure~\ref{fig:hanleIncl} compares the polarization diagrams for the line-center radiation emitted at $\mu=1$ 
obtained with PRD--AD, PRD--AA and CRD calculations, considering
height-independent inclined (i.e., $\theta_B=\pi/4$) magnetic fields of 20\,G with different azimuths $\chi_B$.
These diagrams reveal smaller forward-scattering signals
than those presented in Fig.~\ref{fig:hanle}, due to the smaller horizontal component of the 20\,G magnetic field.
Moreover, both the CRD and PRD--AA diagrams are rotated with respect to the PRD--AD one.
This means that both approximations could lead to a wrong sign of the $Q/I$ and $U/I$ line-center signals.
Figure~\ref{fig:hanleVar_theta} shows a similar polarization diagram,
but for $\chi_B=0$ and different inclinations $\theta_B$ of the 20\,G magnetic field. 
As expected, the $Q/I$ and $U/I$ signals vanish in the presence of a vertical magnetic field, that is $\theta_B=0,\pi$.
As soon as the magnetic field is sufficiently inclined,
%($\theta_B \gtrapprox \pi/12$),
the CRD and PRD--AA calculations
significantly underestimate the amplitude of the line-center fractional linear polarization signals with respect to the PRD--AD modeling.
Interestingly, the CRD diagram
shows always negative line-center $Q/I$ signals.


\section{Discussion}
\label{sec:discussion}

In general, CRD and PRD--AA calculations
significantly underestimate the amplitude of the line-center fractional linear polarization signals,
with respect to the PRD--AD modeling, and even the sign can be wrong.
Providing a simple and intuitive interpretation of these results is, unfortunately, not straightforward. 
Here, we provide a few qualitative insights.
% leaving a more detailed analysis for a future investigation.

% The shape and amplitude of the emissivity profiles resulting from scattering processes
% % descriptions (CRD, PRD--AA, and PRD--AD), as well as the relative differences between them,
% strongly depend on the spectral and angular properties of the radiation field illuminating the atoms.
% The physical origin of the discrepancies between the different scattering descriptions thus lies
% in the approximations that are introduced by CRD and PRD--AA calculations
% in the complex coupling between the frequencies and propagation directions of the incoming and outgoing radiation in the scattering process.
% Indeed, it can be easily verified that in the limit case of a spectrally-flat (and isotropic?)
% radiation field, CRD, PRD--AA, and PRD--AD calculations would be identical.
%
The polarization properties of scattered radiation strongly depend on the detailed spectral and angular dependencies of the incident radiation field.
These dependencies %are fully accounted for only by an
can only be fully taken into account through a
PRD--AD description of scattering processes, while they are necessarily smoothed by the 
averages inherent to the PRD--AA and CRD approximations.
Indeed, it can be shown that the CRD, PRD--AA, and PRD--AD calculations coincide in the limit of a spectrally flat and isotropic radiation field.
It appears reasonable that scattering polarization signals, which are ultimately due to the geometry of the problem and symmetry-breaking effects, 
are enhanced by the PRD--AD approach, which exactly accounts for the complex coupling between the frequencies and propagation directions of the 
incoming and scattered radiation.
Besides, the particular spectral structure and anisotropy
of the solar radiation field depend on the thermal and density stratification of the atmospheric model through non-local RT effects.
Inspecting and predicting these effects in non-academic scenarios is notoriously difficult, and an even more complex task is to
predict how they are impacted by the considered approximations in the modeling of scattering processes.
% \jiri{[But we should also include a note for the referees of SNF's and other projects, not just to admit it is difficult :) Here or later in the conclusions, we could perhaps write something like this:}
%\textcolor{magenta}{However, at the dawn of the new generation of the big solar spectropolarimetric facilities like the Daniel K. Inouye Solar Telescope \citep[DKIST,][]{rimmele20}
%and the future European Solar Telescope \citep[EST,][]{quintero22}, the call to fully include the physical processes affecting these spectra is stronger than ever before.}
In conclusion, it is not possible to clearly identify a single specific reason that explains
the large differences observed between the CRD or \mbox{PRD--AA} calculations with respect to the general PRD--AD modeling.
Nonetheless, these results clearly manifest
the necessity of considering PRD effects in their general AD formulation. 
This fact is even more crucial for full 3D RT calculations, which consider the detailed geometrical structure of the solar plasma.
At the dawn of the new generation of the big solar spectropolarimetric facilities like the Daniel K. Inouye Solar 
Telescope \citep[DKIST,][]{rimmele20} and the future European Solar Telescope \citep[EST,][]{quintero22}, the call to fully include 
the physical processes affecting the polarization of spectral lines is indeed stronger than ever before.

% {It can be shown that the difference between the exact AD expression of the 
% $R^{\scriptscriptstyle \mathrm{II}}$ redistribution matrix and the AA one 
% rapidly increases when the scattering angle (i.e., the angle between the 
% directions of the incident and scattered radiation) approaches either zero 
% (forward scattering) or $\pi$ (backward scattering).
% In a stellar atmosphere, where the radiation propagating outwards is generally 
% more intense than that propagating either horizontally or inwards, forward 
% scattering processes bring a significant contribution to the emissivity in 
% directions close to the vertical. 
% This can explain why the relative differences between the amplitudes of the forward-scattering Hanle effect signals resulting from PRD--AD and PRD--AA (or CRD) calculations significantly increase when approaching the disk center.}

% Finally, we recall that a horizontal magnetic field maximizes the breaking of the axial symmetry and thus the forward-scattering Hanle effect at the disk center.
% Unsurprisingly, the PRD–AD calculations presented Sect.~\ref{sec:horizontalB} provide a forward-scattering Hanle effect signal
% (around 1\,\%), which is about one order of magnitude larger than the ones (around 0.1\,\%) measured in the Ca~{\sc i} 4227 line by \citet{bianda2011fshe}. 
% Clearly, fitting the observations of \citet{bianda2011fshe}
% goes far beyond the scope of this work.
% Indeed, in our 1D calculations, the impact of horizontal inhomogeneities in the solar plasma and the spatial gradients of the plasma bulk velocity
% were not taken into account.
%
A last remark concerns the significant discrepancies between the theoretical calculations presented in Sect.~\ref{sec:horizontalB}, which predict polarization 
signals around 1\,\%, and the observations of \citet{bianda2011fshe}, which show signals one order of magnitude weaker (around 0.1\,\%).
This difference is not surprising, noticing that a horizontal magnetic field
maximizes the breaking of the axial symmetry and thus the forward-scattering 
Hanle effect.
Moreover, the 3D calculations in the CRD limit by \citet{jaume2021CaI} indicate that the presence of an inclined magnetic field 
reduces (and does not enhance) the amplitude of 
the scattering polarization signals produced by both horizontal inhomogeneities of the solar plasma and spatial gradients of the plasma bulk velocity (which are neglected in this work).
The inclusion of mechanisms (i) and (ii) and the use of state-of-the-art 3D MHD atmospheric models are needed to
obtain reliable
information on the magnetic field strength.
We also note that the impact of instrumental effects
% \MyNewEdit{(e.g., optical diffusion and white noise)}
on the observed line-core signals is significant,
especially in forward scattering observations \citep[e.g.,][for the Sr~{\sc i} 4607\,{\AA} line]{zeuner20,delpino21}.

\section{Conclusions}
\label{sec:conclusions}
The results of this work show that a reliable modeling of the fractional linear polarization 
signals produced through the forward-scattering Hanle effect in the Ca~{\sc i} line at 4227\,{\AA}
requires taking PRD effects into account in their general AD formulation.
If the CRD or PRD--AA approximations are considered, the amplitude of the line-center $Q/I$ and $U/I$
signals close to $\mu=1$ could be significantly underestimated, and even the sign can be wrong.
This finding is of clear relevance, especially for the development of new methods
for solar magnetic field diagnostics based on forward-scattering polarization signals
in the Ca~{\sc i} 4227\,{\AA} line.

It can be expected that the results presented here for the Ca~{\sc i} 4227\,{\AA} line generalize to other strong 
resonance lines, for which PRD effects are relevant.
For this reason, it is important to extend this work to other spectral lines of interest for Hanle diagnostics,
such as those recently observed by the CLASP experiments \citep[e.g.,][]{kano2017,rachmeler2022}, 
or those that DKIST (and EST) will allow observing with unprecedented spatial and temporal resolutions.

Finally, %the next step is 
it will be of high interest to consider realistic 3D atmospheric models, which allow including horizontal 
inhomogeneities of the solar plasma and spatial gradients of the bulk velocity.
% Results of
In this respect, the first 3D non-LTE RT calculations, taking scattering polarization and AD PRD effects 
into account, will soon be available thanks to recent software and algorithmic developments \citep[see][]{benedusi2023}.

\begin{acknowledgements}
This work was financed by the Swiss National Science Foundation (SNSF) through grant CRSII5$\_$180238.
T.P.A.'s participation in the publication is part of the Project RYC2021-034006-I, funded by MICIN/AEI/10.13039/501100011033, and the European Union “NextGenerationEU”/RTRP. T.P.A., J.T.B, and E.A.B. acknowledge support from the Agencia Estatal de Investigación del Ministerio de Ciencia, Innovación y Universidades (MCIU/AEI) under grant ``Polarimetric Inference of Magnetic Fields'' and the European Regional Development Fund (ERDF) with reference PID2022-136563NB-I00/10.13039/501100011033.
J.\v{S}. acknowledges the financial support from project \mbox{RVO:67985815} of the Astronomical Institute of the Czech Academy of Sciences.
\end{acknowledgements}

\bibliographystyle{aa}
\bibliography{bibliography}

% \appendix

% \section{Profiles and polarization diagram for a LOS of $\mu=0.9$}

% \MyNewEdit{Valutare .... }

\end{document}